\documentclass[aps,prd,twocolumn,10pt,nofootinbib,showkeys,superscriptaddress]{revtex4-1}
\usepackage{amsmath}
\usepackage{amssymb}
\usepackage{graphicx}
\usepackage{color}
\usepackage{hyperref}

\newcommand{\barr}{\begin{eqnarray}}
\newcommand{\earr}{\end{eqnarray}}

\begin{document}

\title{On the quantum description of the early universe}

\author{Gabriel R. Bengochea}
\email{gabriel@iafe.uba.ar} \affiliation{Instituto de Astronom\'\i
	a y F\'\i sica del Espacio (IAFE), CONICET - Universidad de Buenos Aires, (1428) Buenos Aires, Argentina}

\begin{abstract}
Why is it interesting to try to understand the origin of the universe? Everything we observe today, including our existence, arose from that event. Although we still do not have a theory that allows us to describe the origin itself, the study of the very early era of the universe involves the ideal terrain to analyze the interface between two of today's most successful physical theories, General Relativity and Quantum physics. But it is also an area in which we have a large number of observational data to test our theoretical ideas. Two of the fathers of Quantum physics, Niels Bohr and Werner Heisenberg, shared some thoughts that could be described with these words: \emph{Quantum physics tells us that there is a line between the observed and the observer, and therefore science should be limited to what is observed. We must give up a complete, objective and realistic theory of the world}. This article will orbit around these ideas and summarizes how it is that today, from recent works, we are in a position to try to challenge them (at least in part) through cosmology, seeking the quantum description of the early universe.
\end{abstract}


\maketitle

\section{Introduction}
\label{intro}

The Big Bang model describes the temporal evolution of the universe as a whole. This model has mutated over the decades, to incorporate the results of increasingly precise astronomical observations. In this way, today the model contemplates, in addition to matter constituted by atoms, the existence of cold dark matter (CDM); and for only about 20 years, we think that 70\% of the energy density of the universe is found in something we generically call dark energy, presumably in the form of a cosmological constant $\Lambda$. Also, our current standard $\Lambda$CDM cosmological model includes a phase of rapid expansion at the beginning of the history of the universe called \emph{inflation}. The theoretical pillars of inflation are, fundamentally, General Relativity and the Quantum theory.

The inflationary paradigm is held among the majority of cosmologists as a successful model for addressing the primordial inhomogeneities that represent the seeds of cosmic structure. In fact, the standard prediction from the simplest inflationary model is extremely consistent with recent observations from the Cosmic Microwave Background (CMB) radiation \cite{Planck15}. Then, why think of alternative ideas? Why is the physical mechanism responsible for the generation of the primordial perturbations still a matter of debate?

On the one hand, although we have an excellent explanation, we must remember that throughout the history of science we have had very good scientific explanations for some natural phenomena, which were later inadequate in light of new experiments or theories that allowed the prediction of new phenomena. In some contemporary scientific works we can read phrases such as "The concordance model is now well established" or "there seems little room left for any dramatic revision of this paradigm". But with these statements we could be exaggerating our successes. Therefore, maintaining our critical vision and exploring new ideas are scientifically healthy.

On the other hand, as we mentioned above, inflation is based on a combination of Quantum theory and General Relativity, two theories that are difficult to merge at both the conceptual and technical level. If we want to consider the inflationary account as providing the physical mechanism for the generation of the seeds of structure, such account must contain an explanation for some recently staged problems, e.g. \cite{Ijjas,Penrose2017, vafa19}, as well as give a satisfactory answer to the following question: why does the quantum state that describes our actual universe not possess the same symmetries as the early quantum state of the universe, which happened to be perfectly symmetric?

Since there is nothing in the dynamical evolution (as given by the standard inflationary approach) of the quantum state that can break symmetries, the traditional inflationary paradigm is incomplete in that sense. As we will see below, this is closely related to what is known as \emph{the measurement problem} in Quantum physics \cite{Wigner63, Omnes, Maudlin95,Becker}, and which is notoriously exposed in the case of the quantum description of the very early era of the universe.

There are promising alternatives to standard formalism today that allow, on the one hand, to accommodate empirical evidence, and on the other, to construct an objective and complete image of the world. In order to evaluate and classify the possible alternatives to achieve this, Tim Maudlin stated the measurement problem in a formal and general way, showing that there are three statements that are mutually inconsistent \cite{Maudlin95}. In short: A) the physical description provided by the quantum state is
complete, B) quantum states always evolve according to the Schr\"odinger equation, and C) measurements always have definite results. And in such a work, the author concludes that any real solution will demand new physics and that, in particular, the so-called collapse theories and hidden variables theories have a good chance of succeeding.

In this article, with a pedagogical approach aimed at science students, teachers and also non-expert colleagues, we will make a description of the quantum problems that must be faced when it comes to giving a description of the emergence of seeds of structures in the early universe. Throughout the manuscript, we will mention various approaches to these problems and, following Maudlin's conclusions in \cite{Maudlin95}, we will emphasize the proposals known as \emph{objective collapse theories}.

In Sect. \ref{secuno} we will highlight some differences between classical and quantum physics; in Sect. \ref{secdos} we will describe the measurement problem in Quantum physics; in Sect. \ref{sectres} we will address the cosmological case; in Sect. \ref{seccuatro} we will mention an approach that seeks to solve the aforementioned problem; and finally, in Sect. \ref{conclusions}, we will present some conclusions.

\section{Classical vs. Quantum Physics}
\label{secuno}
Here we are going to refer to \emph{classical physics} as that described by Newton's laws (or by Einstein's theories of Relativity). We use these laws to calculate and predict, for example, what are the values ​​of the position and velocity of an object at a given time. Given the values ​​in an instant, Newton's laws allow us to perfectly predict its \emph{trajectory} in space. From this point of view, classical physics is objective, complete and realistic. Briefly, with \emph{objective} we mean that it does not depend on someone making the measurements (it does not need an observer); it is \emph{complete} because in the theory there is all the information necessary to describe the properties of objects (that is, every element of ‘reality’ has a counterpart in theory); and \emph{realistic} because the elements of the theory really describe \emph{real} objects that have properties with well-defined values. Those objects exist in the world regardless of someone observing them and, with the theory, one can predict those values.

On the other hand, in standard Quantum physics, physical properties such as the position or velocity of an object in general \emph{do not have defined values until a measurement is carried out}\footnote{By 'standard' Quantum physics we are referring to the so-called \emph{Copenhagen interpretation}, which is adopted by the vast majority of authors in textbooks. However, the various interpretations of Quantum physics studied at present face the problems mentioned in this article. See for example \cite{casta17}.}. All the accessible information of a quantum system is contained in what we call its \emph{wave function}. This function is not something that one can observe, but it is what allows us to calculate probabilities, with a rule for that purpose given by Max Born in 1926, which constitutes one of the postulates of Quantum Mechanics. Probabilities for what? For the possible values of the physical quantities that could be obtained\footnote{We will use the terms 'physical quantities' and 'physical properties' of objects as synonyms.} (such as the position, for example), if we made a measurement with some appropriate device to measure the physical property that we are interested in knowing (the position of the object in our example). With this theory we have been able to describe in an extremely precise and successful manner numerous phenomena and experiments: from atoms and elementary particles, to how the Sun and the other stars shine, nuclear energy, lasers and all the electronics we use in our daily lives, to mention just a few examples. In fact, our idea is that the whole universe in its essence is quantum and then our daily macroscopic theories would be just very good classical approaches to something deeper and more fundamental. But how is it that the macroscopic objects of our daily lives, being composed of atoms, do not seem to be described by the physics that so successfully describes atoms?

In 1927, Werner Heisenberg proposed what is known as the \emph{Uncertainty Principle}. This principle tells us that the better determined is the value of a certain physical quantity in a certain quantum state (the position, for example), the less determined will be the value of another \emph{conjugate} quantity (its momentum, or velocity). Recall that, according to Newton's classical physics, objects have, at any time, all the values ​​of all properties perfectly defined. On the other hand, quantum uncertainties, together with the Born probability rule, give us the range in which the property values ​​are most likely to be if we made measurements. Until we make measurements, with devices designed to know the values ​​of observable physical quantities, these (and even the properties themselves) are not determined and they are not independent. In this way, although we measure some properties, others will remain undefined or will be altered. Then, the most general quantum state will be a state of superposition. By superposition we mean that, as the values ​​of some properties are not determined, the quantum state is a "combination" of the possible states and the Born's rule allows us to calculate, from the superposition, the probabilities of the possible values.

Here is where the best-known pet in physics comes into play: Schr\"odinger's cat. Erwin Schr\"odinger was the one who managed to formulate in 1925, following the ideas of Louis de Broglie, an equation (today known as \emph{Schr\"odinger equation}), which determines how the wave function of a quantum system and its probabilities evolve over time. It is the pillar equation of Quantum physics. And with it we will raise what is known as the paradox of Schr\"odinger's cat.

\section{The measurement problem in quantum physics}
\label{secdos}

The theoretical experiment that Schr\"odinger thought in 1935 consists of the following: inside a closed box without windows there is a cat. In the box next to it is a bottle that contains a deadly poison and there is also a \emph{random} atomic device with two possible states, with a 50\% probability each. One of the states of the device has a 50\% probability of acting on a hammer breaking the bottle, releasing the poison and thus killing the cat, at some time that we cannot know with precision. The other state has a 50\% chance of not acting, and therefore the cat will remain alive. But, and here comes the important point, Quantum physics tells us that the most general state of the atomic device is a combination of the two possible states. But both, the device and the bottle with the poison, the hammer, the cat and the box are made of atoms. Therefore, everything should be described by Quantum physics, if this, as it is supposed, is applicable to everything in the universe. If the atomic device is initially in a quantum state of superposition, considering both the apparatus and the cat as quantum systems that interact with each other, the state of the cat will \emph{get entangled} with that of the device, and then, it will also be in a state of superposition until some measurement is made. If we wanted to know, for example, something about the "\emph{liveliness}" property, according to Quantum physics (in its standard interpretation), until we make a measurement of that property the most general quantum state is a superposition of the two possible states: \emph{alive-cat} and \emph{dead-cat}, with 50\% probability for each possibility. That is, the cat is not alive or dead. There is no definite value of the \emph{liveliness} property\footnote{But be careful: it is not that it could already have a value but we do not know it because of our ignorance. It has no defined value yet, until a measurement is made. And when we measure, there are still many other properties that cannot have their values defined simultaneously.}. And it is a perfectly valid and possible state for Quantum physics.

The Schr\"odinger equation, which allows us to know the evolution in time of the state of any quantum system, determines that the cat (or our knowledge about the liveliness of the cat in the standard interpretation) will remain in the "\emph{alive-dead}" superposition state until someone or some device for this purpose makes a measurement (open the box, for example). Schr\"odinger's equation does not destroy neither superpositions nor probabilities and does not break symmetries; it is deterministic and reversible. With determinist we mean that you can know perfectly at every moment what the wave function of the system is, and reversible because at all times we can calculate backward or forward in time what the value of the wave function is. We will call this "\emph{Process A}".

But after a measurement, something happens. The wave function "\emph{collapses}" and a well-determined value is obtained (for example, the life of the cat results in \emph{alive-cat}). This other process is random (it could have been \emph{dead-cat}), it is irreversible (once we measure, we cannot know if before that the cat was alive, dead, or alive-dead) and, therefore, some information is lost. We will call this second process "\emph{Process B}".\footnote{These processes are referred to as U process and R process respectively in \cite{PenroseB} and as Process 2 and Process 1 respectively in Everett's seminal paper \cite{Everett}.}

Similarly, when a scientist prepares a system in a laboratory (particles in an accelerator, for example) in a state of superposition (for example, for the position) and then that system interacts with some appropriate measuring device to measure the position, the states of the indicators and the needles of the apparatus will become entangled with those of the system and, then, the whole set (system $+$ apparatus) ends in a state of quantum superposition. While nothing or no one makes a measurement, the needles of the device would continue in a state of superposition. However, of course, this is never observed in the laboratory.

Then, if Quantum theory is applicable to everything, why small objects such as atoms can remain in states of superposition, but everyday objects, such as my chair or the needles of a device, are not in a superposition of two places at the same time?

The general situation is, then, that until we make a measurement, the most general state of a physical system is to be in a superposition of states, and quantum uncertainties, together with the Born's rule, tell us the ranges of possible and more likely values of the properties. And then, when we carry out a measurement to know some physical magnitude, the \emph{$X$} position, say, the wave function collapses and a well defined value is obtained for \emph{$X$}, compatible with the Uncertainty Principle.

But how does a system go from a superposition of states for \emph{$X$} to another state without superpositions, and with a well-defined value of \emph{$X$}, if the Schr\"odinger equation does not destroy superpositions? If someone (or something for that purpose) made a measurement, it would reveal to us in what state the system is. But something \emph{external} should cause the wave function to collapse to another well-defined state. On the other hand, it is important to say here that, in addition, the concept "\emph{measurement}" is not satisfactorily defined within Quantum physics. How large must an object be so that its state collapses and is not in a superposition? About the size of a cat? When does a measurement happen? Quantum theory does not tell us. There is no clear criterion of when we should use the evolution given by \emph{Process A} and when to use \emph{Process B} that determines the collapse of the quantum wave function. This is known as "\emph{the measurement problem}" in Quantum physics, which can be stated in a formal manner as, for instance, Maudlin did \cite{Maudlin95} and as we already mentioned in the Introduction\footnote{For more details see, for instance, Refs. \cite{Becker,okon14}. Some people choose to deny the existence of this problem, stating that Quantum physics is only about calculations to predict probabilities and that when we make measurements in the laboratory everything fits perfectly. But we will see in "the cosmological case" that this position cannot be sustained in a completely satisfactory manner.}. In classical physics, things happen according to certain laws, no matter if there are observers who decide when and how to make measurements so that one or the other law of evolution is applied. Why does the quantum realm seem to be so different?

We have said that quantum evolution, dictated by the Schr\"odinger equation, cannot produce the collapse of the wave function. So what produces it? There are many proposals that try to answer, from various perspectives, this question. We will mention some of them here.

Some scientists, as Bohr did, argue that physics should take care only of what is observed. That is, giving up an objective theory, free of a description of the world by whom it decides to observe. Others say it is the fault of the measuring device. The device interacts with the object, \emph{Process B} is triggered changing the state and the collapse occurs. But how large must an apparatus be to act as an apparatus? Is an electron orbiting an atomic nucleus measuring the protons of the nucleus? Is it perhaps the observer who causes the wave function to collapse? And what does an observer represent? A human? A chimpanzee? A cat? These proposals are the best known of those that deny Maudlin's statement B).

Other authors argue that although the evolution of quantum states is given at all times by the Schr\"odinger equation, the result obtained by an experimenter when making a measurement is not the only one. Such is the case of the \emph {many worlds} approach (based on the original idea of H. Everett \cite{Everett}) where, once the measurement is carried out, something happens in such a way that all the possible results are obtained in (real or not) a diversity of universes\footnote{To be fair, in his seminal paper Everett only referred to '\emph{relative states}'}. Therefore, a state of superposition is nothing other than the promise of the existence of other worlds.

Another well-known approach proposes that since an object completely isolated from the rest of the world does not exist, \emph{the environment} interacts with the object, alters its state causing the macroscopic superposition of all possible states to disappear, triggering a sort of "effective collapse", and thus resolving the whole problem\footnote{Although this approach (known as \emph{quantum decoherence}) in some cases manages to partially solve the problem, it does not end up being a satisfactory solution and also usually requires an external observer to subjectively decide issues or carry out measurements. A detailed analysis of these and other problems of this approach that we are not mentioning here can be seen in \cite{okon16, Adler01}.}. But what is the rule to apply to decide in each case where the object ends and where the environment begins and ends? What or who decides what is and what is not environment? We are? So the quantum nature of the world depends on our existence?\footnote{One might argue that if Quantum physics is a description of nature, it is reasonable to think that it would depend on the existence of its descriptors. The measurement problem is precisely that something outside the standard Quantum theory is needed to solve it. For instance, observers or descriptors. But in the cosmological case, to explain the early times of the universe, we will see that it will be difficult to feel comfortable with this approach.} This proposal and the Everettian interpretations are some of the approaches that somehow discard Maudlin's statement C).

The reality is that none of this is well defined in Quantum theory and none of this has been able to completely solve the measurement problem. So, the question \emph{how does a quantum system move from a state of quantum superpositions to another state without superpositions?}, to this day it does not have a complete and satisfactory answer.

Why then is Quantum physics so successful if it has this measurement problem? The answer is that Quantum physics is about making measurements, and when we want to use the theory, in practice, dividing the world between the observed and the observer is easy in a laboratory even though the theory does not provide us with a clear rule. In general, the separation between what is the object of study and what constitutes the apparatus is very well defined. At the most, it will be enough to incorporate more components to the quantum system until the predictions are no longer altered, and thus the results will be consistent with the observed. On the other hand, the aforementioned separation in laboratory situations is always simple, because the scale of the quantum systems of study (atoms, for example) is very far from the human scale, from the scale of the devices and also from the resolution and precision of our devices.

But this cannot be entirely satisfactory. Hartle, for instance, mentions that the usual formulations of Quantum mechanics are inadequate for cosmology, since these formulations assumed a division of the universe into "observer" and "observed" and that fundamentally quantum theory is about the results of measurements. But measurements and observers cannot be fundamental notions in a theory which seeks to describe the early universe where neither existed \cite{Hartle93}.

And here is when we move to the realm of the universe on large scales. The problem of quantum measurement worsens terribly in the cosmological case\footnote{One of the first references where this was noted is in the Introduction of one of J. Bell's works \cite{Bell81}.}. Let's see why.

\section{The cosmological case}
\label{sectres}

The measurement problem, in the cosmological context, is a subject that has received much less attention from the physics community. However, we should point out that some researchers in the field, such as Hartle and Penrose, have pointed out the need to generalize quantum mechanics to deal with cosmology \cite{Hartle93,Hartle97,Hartle06,PenroseB,Penrose96}. The proposal for generalization of quantum physics using a scheme based on the realms of decoherent coarse-grained histories proposed by Hartle is an example, but we will not discuss it here since it exceeds the scope of this article.

As we mentioned in the Introduction, the Big Bang model, with which we seek to describe how the origin of the universe and its temporal evolution were to this day, fundamentally involves the two pillars of modern physics: Gravitation (Einstein's theory of General Relativity) and Quantum theory.

And more precisely, when we want to understand how the first moments of the universe were and how the first "seeds" (the \emph{primordial inhomogeneities}\footnote{Technically, these seeds of structure or "inhomogeneities" are called \emph{cosmological perturbations}. Therefore, we will use the terms inhomogeneities or perturbations interchangeably.}) of the cosmic structure emerged (and which then ended up in, say, galaxies), Quantum physics takes an extremely leading role in this description. These first moments of the universe are described by a model we call \emph{cosmic inflation}.

Fundamentally with the work of Alan Guth in 1981 \cite{Guth81}, and by works of Andrei Linde, Paul Steinhardt, Andreas Albrecht, Viacheslav Mukhanov, Alexei Starobinsky and Stephen Hawking among others \cite{Linde81,albrecht82,Mukhanov81,Starobinsky79,Bardeen83,Brandenberger84,Hawking82}, the proposal arose that if at the beginning from its history ($\sim10^{-35}$ seconds) the universe had gone through a brief inflationary phase of accelerated expansion driven by an exotic field called \emph{inflaton}\footnote{Many physical phenomena of nature are described using fields. Such as the electric field, the magnetic field, the gravitational field, etc. The \emph{Inflaton} is an exotic scalar field, whose potential energy would have been dominant only at the beginning of the universe causing its expansion to be accelerated.}, some problems then known from the standard hot Big Bang model could be resolved and all of them with the same mechanism. We will not go into detail here about what those problems were, since it is not the aim of this article.

From a scientific meeting held in Cambridge, UK, in 1982 (the Nuffield Workshop organized by Gibbons and Hawking), and with the ideas of a 1965 Andrei Sakharov work in mind \cite{Sakharov65}, the mentioned authors began to show that the emergence of the seeds of the structures in the universe could have occurred due to "\emph{quantum fluctuations}"\footnote{Below it will be clear what we mean by this concept.} of the inflaton field during that same inflationary process. The gravitational evolution of those seeds generated in inflation, with the passage of time, would have ended in everything we observe today in the sky; and that evolution, in addition, seems to be very well reproduced with numerical simulations that are carried out with large computer arrangements.

One of the observational lines that has had more development and has achieved more data in recent decades, is the one that deals with the analysis of what is known as the \emph{Cosmic Microwave Background} (CMB) radiation. This cosmic background is electromagnetic radiation that reaches us with a practically identical spectrum from all directions of the sky (today with greater intensity in the microwave range), and characterized with an average temperature of only about 2.7 K. The existence of this radiation was predicted in the late 1940s by George Gamow and others, but was discovered in 1965 by Arno Penzias and Robert Wilson. We think that it comes from the time when the first neutral atoms in the universe were generated, about 380 thousand years after the Big Bang. The statistical analysis of the small differences in the temperature of this radiation that are observed in the different directions of the sky constitutes the study of what is called the \emph{anisotropies} of the CMB. These very small temperature differences are one part in one hundred thousand. Theoretically, as the authors mentioned above began to show, we expect these tiny temperature differences to be present in the sky, since they would be the result of the evolution of the seeds (\emph{primordial perturbations}) generated at the beginning of the universe, and whose origin we attribute it to the inflation mechanism. The surprising fact is that the anisotropies observed in the sky are exactly like those predicted by the inflationary model, and without this model, today it would be quite difficult to explain the origin of what we observe\footnote{Some recognized authors such as P. Steinhardt, R. Penrose, R. Brandenberger and others have been stressing that inflation has some serious problems, see for instance \cite{Ijjas,Penrose2017, vafa19}. And it is fair to mention that there are some variants and alternatives to the inflationary paradigm, including in that list, for example, models of cyclic universes. But to date they have not been able to be sufficiently competitive.}.

Then, here we have this situation: we observe large structures (galaxies and clusters of galaxies) and also small anisotropies in the temperature of the cosmic microwave background. We assume that its origin dates back to the beginning of the universe, where the original cosmic seeds must have existed. We do our calculations and everything fits perfectly between theory and observation. But where did those initial seeds come from? How were they generated during cosmic inflation?

This is where our main protagonist of the article reappears: Quantum physics.

How do we apply the Schr\"odinger equation of Quantum physics to the case of the inflaton at the beginning of the universe? What do we think was the initial quantum state of the primordial perturbations with which we make our calculations to make theoretical predictions?

At the moment when the inflationary phase begins to occur, we have, on the one hand, the spacetime (whose evolution is described by Einstein's equations of General Relativity) and, on the other hand, the inflaton field dominating the energy budget of the early universe, producing the accelerated expansion, and whose quantum inhomogeneities we want to know how they emerged\footnote{Although the most standard version proceeds by quantizing both spacetime and the inflaton field, here we will adopt the approach that spacetime (at least since inflation) is always classic and that quantization is done only to the inflaton field. This does not change at all the central point of this article, the problems that here are addressed and the conclusions.}. Then, Einstein's equations tell us how spacetime (its curvature) reacts and is affected by the presence of the inhomogeneities of the inflaton field.

We assume that far back in time, at the beginning of inflation, the spacetime was the most symmetrical and simple of all. It was isotropic (there was no privileged direction) and homogeneous (there was no privileged point or place in space)\footnote{Before inflation occurs, spacetime may have been highly inhomogeneous (as a product of physics that we do not yet know fully and satisfactorily). The standard argument is that, once the accelerated expansion that leads the universe to a Inflationary phase starts, this produces that any inhomogeneity is suppressed exponentially. Therefore, at the beginning of inflation, spacetime is typically assumed to be isotropic and homogeneous and then the anisotropies observed today in the CMB are thought of as the exclusive result of the inflationary process.}. We also assume that the inhomogeneities of the inflaton field were, at that same time, in a quantum vacuum state perfectly isotropic and homogeneous. That is, a state with definite energy and which also had the same symmetries as the initial spacetime\footnote{A vacuum state is one that, at least for some instant of time, has a well-defined energy and is generally minimal. While at this point there is a technical problem that we will not address here, which has to do with the fact that there is no single manner to choose a quantum vacuum state in an expanding universe, the consensus is that the initial vacuum state for cosmological perturbations was what is known as the \emph{Bunch-Davies vacuum}, which is perfectly isotropic and homogeneous.}. We could start from a different initial situation, a little more complex, without some symmetries, or that already contains the cosmic seeds of future galaxies beforehand. But then we would find the extra task of developing another theory to explain why the universe was born with a more complex situation and not the simplest.

As with any quantum system, we can now calculate the expected values and quantum uncertainties of perturbations in the quantum vacuum state. And, in the same way as when we said that in a laboratory experiment, until a measurement does not occur for the position of a particle in general it is not defined, that it is in a state of superposition, and that the quantum uncertainty tells us in what range of possible values we can find most likely when we make a measurement, the same should now apply to our case of the quantum universe. In the case of the laboratory, when we measure some physical property the wave function collapses, and then our devices give us defined values.

It is, then, when the central question of this article arises: how do we arrive at an anisotropic and inhomogeneous quantum state (with the seeds of structures), from a vacuum state, with superpositions, perfectly isotropic and homogeneous (without cosmic seeds)? We have said that the quantum state of a system contains all the information of that system, and that the evolution of any quantum state is dictated by the Schr\"odinger equation, which does not break any symmetry or destroy quantum superpositions. Until the symmetries are broken and the quantum state changes, the space will remain isotropic and homogeneous, the curvature of the space will be the same at each point and, therefore, there will be no chance of a galaxy or anything else appearing in the future.

Who or what made a measurement producing the collapse, the loss of the initial symmetries and the emergence of the seeds of structure at the beginning of the universe, giving non-null and well-defined values for the perturbations of the inflaton and spacetime? Was it any device? Any observer? The environment? Of course, we want to think that none of this existed at the beginning of the universe\footnote{For a discussion regarding that "an environment" cannot solve the problem, see for example Sect. 3.2.1 of Ref. \cite{okon16}. Works based on decoherence \cite{kiefer09,halliwell,kiefer2,polarski} led to a partial understanding of the issue. Nevertheless, this argument by itself cannot address the fact that a single (classical) outcome emerges from the quantum theory. In other words, decoherence cannot solve the quantum measurement problem \cite{Adler01, schlosshauer}, a complication that, within the cosmological context, is amplified due to the impossibility of recurring to the "for all practical purposes" argument in the familiar laboratory situation. Other cosmologists seem to adopt the Everett "many-worlds" interpretation of quantum mechanics plus the decoherence process when confronted with the quantum-to-classical transition in the inflationary universe, e.g. \cite{mukhanov2005}. Regarding this point, we would like to refer the reader to other Refs. \cite{Sudarsky11,kent,stapp} where arguments against decoherence and the Everett interpretation are also presented.}.

Typically, the most orthodox version of this analysis draws on the Uncertainty Principle to say that the initial \emph{"quantum vacuum fluctuations"}\footnote{Note that the correct thing would be to talk about the \emph{quantum fluctuations of the inflaton field} in the vacuum state. In fact, this lightness in the discourse is often accompanied by phrases such as "\emph{the quantum fluctuations of the vacuum energy}", which is totally wrong since the quantum uncertainty of the energy in the vacuum state is exactly zero. Some arguments involving quantum vacuum fluctuations as a mechanism for solving the so-called cosmological constant problem also proved inadequate \cite{Bengochea20}.} are the mechanism for generating the seeds of the structures. From this approach, quantum fluctuations have real existence in the universe. That is, quantum fields acquire real, random, but well-defined values at every time, and make the curvature of spacetime change (and oscillate like a spring, for example), in the same way as in Newton's theory the position of a tennis ball is taking defined values following a trajectory in space. This contradicts what we understand of standard Quantum physics, and is not what we have in mind when experimenters do their job in a terrestrial laboratory (they make measurements!). Quantum fluctuations are nothing other than quantum uncertainties\footnote{The word "\emph{fluctuations}" in physics is often used (and confused) in several different contexts. It can mean the variations or the range of values for some characteristic of objects within a set (variations in the height of a set of chairs, for example); or it can also refer to variations in different regions of something homogeneous (such as waves in the sea); or, as in this article, it can also refer to quantum uncertainties.}. And a quantum uncertainty other than zero for the perturbations in the vacuum state, the only thing that gives us, together with the Born's rule, is the range of its most probable values, but that \emph{there are no defined values} for the perturbations until a measurement is carried out. As in a laboratory, we must always talk about possible measurement results so that Quantum physics predictions make some sense. Therefore, under this analysis approach, all points of space must remain equivalent, space remains isotropic and homogeneous, and there are no seeds of structure of any kind. Quantum vacuum fluctuations cannot be the seeds to form structures. The inflaton field in its vacuum state has fluctuations (quantum uncertainties) but there are no inhomogeneities \cite{Sudarsky11}.

The standard approach, then, cannot fully justify how the initial perturbations appear in the early universe. It requires some process that acts "\emph{as a measurement}", as in the laboratory, and produces something like a collapse of the wave function changing the quantum state. This new state must contain the perturbations or seeds of the cosmic structures. In the next section, we will analyze one of the current proposals that aims to address this issue.

\section{Facing the problem and other related issues}
\label{seccuatro}

One of the approaches that seeks to address the aforementioned problem [removing the Maudlin's statement B)] has, as its central idea, the proposal that in order to solve the measurement problem in Quantum physics, non-standard quantum theories should be explored. Theories where the collapse of the wave function is self-induced by some novel mechanism. Known as models or \emph{objective collapse theories}, they are an approach different from those mentioned above in Sect. \ref{secdos}, and are currently of particular interest in the case of the quantum origin of the primordial seeds of the structures. We will describe in this section some details about these ideas.

From the mid-1970s and more intensely in the 1980s and 1990s, authors such as Pearle, Ghirardi, Rimini, Weber, Penrose, Diosi and others \cite{Pearle76,Pearle89,Ghirardi86,Penrose96,Diosi87,Diosi89} began to seek and develop modifications to the Schr\"odinger equation to alter the evolution of the quantum state and that the collapse of the wave function occurs, without external observers or devices present that have to make measurements; and in that way, solve the measurement problem in Quantum physics. The main idea is that, with the same theory, microscopic phenomena (excellently described by standard Quantum theory) as well as macroscopic phenomena that do not show superpositions, can be explained (in these theories, Schr\"odinger's cat is alive or already dead before we open the box). That is, they sought to achieve a theory that, with the same equation of evolution, can be described states of superpositions of electrons, for example, but also that it can explain why cats and everyday objects are not in superpositions.

The modifications to the Schr\"odinger equation must be such that quantum superpositions for macroscopic objects disappear and locate them in space, in the way we see what happens in our daily lives. To do this, the equation must incorporate some "\emph{amplification mechanism}" that discriminates small objects from large, and that the dynamics itself causes the collapse and leads any initial quantum state to another, stochastically (to explain the randomness observed in the results of laboratory measurements), and reproducing the successful predictions of the quantum probability rule proposed by Max Born. Detailed reviews can be found, for instance, in \cite{Bassi1, Bassi2}.

Guided primarily by the ideas of Diosi and Penrose, in 2006 Sudarsky and collaborators proposed applying the ideas of modifying the Quantum theory to the cosmological case \cite{PSS06}. That is to say, to incorporate in the Einstein's equations for the dynamics of the universe the effects of the self-induced collapses of modified quantum theories. Thus, during the period of cosmic inflation, there would have been spontaneous collapses in the initial quantum vacuum states, similar to a measurement, so that the final result is a new quantum state with different symmetries than the initial ones, without quantum superpositions, turning on the perturbations and giving them non-zero defined values, altering the curvature of spacetime, and thus creating the seeds of structure in the universe. Without observers or measuring devices.

With these modifications, theoretical predictions can be made, which then allow these theories to be tested and thus be able to say something about their viability to explain the precise observations, for example, of the CMB. Some predictions have proved very interesting since they have been able to explain certain observational constraints in a more natural and clear way than in the standard case (see for instance, \cite{Daniel10,Tejedor12,Tejedor12B,Pedro13,Bengochea15,Leon16,Landau12,Picci19,Mariani16,Bouncing16,ModosB})\footnote{Other authors have explored similar ideas and some of these works can be seen, for instance, in \cite{Martin12,Das13,Ellis18}. There is also another approach, which denies Maudlin's statement A) mentioned in the Introduction, where it is argued that the quantum state does not contain all the information necessary for the description of a quantum system. In this way, the addition of \emph{hidden variables} and the equations that determine their evolution is required. The best known case is the \emph{de Broglie-Bohm model} \cite{bohm}. Applications to the cosmological case can be seen, for example, in \cite{Neto12,goldstein15,Neto18}}.

These ideas continue to evolve. More recently, it has been shown that this approach would allow addressing other questions of gravitational origin that have been open for many years. Such are the cases of the information paradox in black holes and the origin of dark energy \cite{Modak14,Modak15,Modak16,Josset17}. The proposal of some authors that other universes besides ours could exist, is tied, in part, to the occurrence of the inflationary phase at the beginning of the universe and to the theoretical problems mentioned above. Therefore, this approach could also make the possibility of the so-called \emph{multiverse} a myth \cite{Leon17,multiverso}.

Modified quantum theories are not yet in their final versions, they face their own questions and problems and are a challenging work in progress. To mention just a few of them, the origin and nature of the stochastic noise contained in some versions of these theories are unknown (some people think that its origin could be gravitational \cite{Diosi84,Diosi87,Diosi89, Penrose96}); best known applications are still nonrelativistic (a relativistic model under exploration can be found in \cite{Pearle2015}), and collapse process appears to violate some conservation laws. For example, particles gain energy from the narrowing of wave functions by collapse. Recently, some authors explored the status of conservation laws in classical and quantum physics. They found that in some contexts conservation laws to be useful, but often not essential \cite{Maudlin20}. If this turns out this way, it can be used to find, for example, a possible origin of dark energy \cite{Perez19, Josset17}. A technical analysis of the various problems that collapse theories face can be found for example in \cite{Pearle06A,Pearle06B}.

\section{Conclusions}
\label{conclusions}

The Big Bang, our model to describe the evolution of the universe, fundamentally combines two of the most successful theories developed in the twentieth century: General Relativity and Quantum physics. The model successfully describes and explains numerous cosmological observations. Even so, we know that it cannot be the final version of the story. An extrapolation of this model to the very origin of the universe is not entirely justified, and could even result in too simplistic and daring. To this day, we still do not have a fully satisfactory quantum theory of gravity that manages to unify both theories. So we do not know, among other things, the origin and nature of spacetime, nor the origin of quantum fields as the case of the inflaton.

There are several proposals that try to respond, from various perspectives, to the problems mentioned in this article. Today all options have their advantages and their own open problems. Within these proposals we have focused particularly on those known as objective collapse theories, which seek to achieve a Quantum theory that somehow is (in some sense) realistic, complete and objective that challenge the thoughts of renowned scientists like Bohr or Heisenberg. These theories are one of the current candidates under study with which not only could the measurement problem in Quantum physics be solved, but the quantum origin of structures in the early universe could also be explained in a more complete and clear way.

Some quantum secrets have not yet been revealed: could in the future the same mechanism be able to solve the quantum measurement problem and, at the same time, other gravitational problems that still have no satisfactory solutions? This approach, perhaps, could also serve as a guide in the search for a quantum theory of gravity.


\begin{acknowledgements}
I thank Daniel Sudarsky, Gabriel Leon and Leo Vanni for productive conversations about the various topics addressed in this article, and the anonymous referee for their valuable suggestions. Also, I acknowledges support from grant PIP 112-2017-0100220CO of CONICET (Argentina).

\end{acknowledgements}
\bibliography{bibliografia}
\bibliographystyle{apsrev}
\end{document}